\documentclass{imamat}

\usepackage{amsmath}
\usepackage{amsfonts}
\usepackage{amssymb}

\newcommand{\ue}{\mathrm{e}}

\newcommand{\sgn}{\mathrm{sign}}
\newcommand{\ri}{\mathrm{i}}
\newcommand{\eps}{\varepsilon}

\begin{document}

\title{Integrable reduction and solitons of the Fokas-Lenells equation}

\author{\name{Theodoros P. Horikis$^*$}
\address{Department of Mathematics, University of Ioannina, Ioannina 45110, Greece
\email{$^*$Corresponding author: horikis@uoi.gr}}}

\maketitle

\begin{abstract}
{Novel soliton structures are constructed for the Fokas-Lenells equation. In so doing,
and after discussing the stability of continuous waves, a multiple scales perturbation
theory is used to reduce the equation to a Korteweg-de Vries system whose relative
soliton solution gives rise to intricate (and rather unexpected) solutions to the
original system. Both the focusing and defocusing equations are considered and it is
found that dark solitons may exist in both cases while in the focusing case antidark
solitons are also possible. These findings are quite surprising as the relative
nonlinear Schr\"odinger equation does not exhibit these solutions.}

{Fokas-Lenells equation, KdV equation, dark and antidark solitons, multiple scales,
modulation instability.}
\\
2020 Math Subject Classification: 35C05, 35C08, 35C20, 35Q51, 35Q55, 35Q60.
\end{abstract}

\section{Introduction}

The theory of multiple scales analysis has been an invaluable tool in the study of
physical phenomena and the underlying equations that describe them. Most of these
systems, in their original form, are very difficult or even impossible to study
analytically and in some cases even numerically. By the use of multiple scales theory,
however, they can be reduced to more manageable systems whose properties are also quite
remarkable. Prime examples are the Euler equations in water waves and Maxwell's equations
in electromagnetics; the first may be reduced to the Korteweg-de Vries (KdV) and
nonlinear Schr\"odinger (NLS) equations for shallow and deep water waves \cite{waves},
respectively, while the latter to the NLS equation under quasi-monochromatic
approximation in optics \cite{waves,kivshar_book}.

These two equations (often referred to as {\em Universal} due to the numerous
applications they appear in) spawned a new field and direction in the study of nonlinear
partial differential equations, namely integrable systems. These systems exhibit
remarkable properties which can be systematically derived using the Inverse Scattering
Transform (IST) \cite{ist} and the newly introduced extension, often referred to as the
Unified Transform or Fokas method \cite{fokas1}.

Besides the obvious interest in integrable equations and their individual properties
another important direction is the connection between them (or the reduction of one to
the other). The Miura map \cite{miura} provides such a connection between the KdV and
modified KdV equations, both of which are integrable. As this is not always possible,
namely finding an explicit, exact way to transform one equation to another the method of
multiple scales and asymptotic analysis has been employed to connect integrable systems
\cite{zakharov,horikis2}.

The purpose of this work is to provide an asymptotic connection/reduction of another NLS
type integrable system to the KdV equation. We are refereing to the Fokas-Lenells (FL)
equation \cite{fokas2,lenells3,lenells4} which is an integrable generalization of the NLS
equation, derived to model nonlinear pulse propagation in monomode optical fibers when
certain higher-order nonlinear effects are taken into account \cite{lenells4}.
Mathematically, this equation is related to the NLS equation in the same way that the
Camassa-Holm equation is related to the KdV equation \cite{lenells5,lenells3} and its
soliton solutions obtained using different methods \cite{lenells5,lenells6,veks,ai}, the
long time asymptotics for the system have been discussed in Ref. \cite{xu} while its
nonholonomic deformation in Ref. \cite{kundu} and rogue type solutions in Ref.
\cite{yang,ye,geng}. A variable coefficient system has also been studied in Refs.
\cite{lu,wang1,wang} and a nonlocal variant in Ref. \cite{zhang}.

As such, starting from the FL system we use a multiple scales scheme that allows us to
reduce the system to a KdV equation whose soliton solutions will be later used to
describe solitons of the original system. Both the focusing and defocusing cases are
considered. What is rather surprising is that in the focusing case (expected to be
unstable as is the relative NLS limit) there is a region of stability which allows for
two types of solitons to exist: dark and antidark intensity dips/humps off of a stable
background. These are unique to the FL system and hence the focusing case provides a
singular limit which does not fall back to the NLS case.

\section{Stability and multiplescale analysis}

To begin our analysis, consider the FL system \cite{fokas1,lenells3,lenells4}
\begin{equation}
\ri u_t -\nu u_{tx}+\gamma u_{xx}+\sigma|u|^2(u+\ri \nu u_x)=0
\label{fl}
\end{equation}
where $\gamma$, $\nu$ are real constants and $\sigma=\pm 1$. In the case of the regular
NLS equation ($\nu=0$), the case $\sgn(\gamma\sigma)=1$ corresponds to the focusing case
and admits bright soliton solutions (decaying to zero at infinities), while the case
$\sgn(\gamma\sigma)=-1$ is the defocusing case where dark solitons (which tend to a
constant background at infinity) exist. The same terminology is used here.

Furthermore, in the focusing case the NLS equation is modulationally unstable. That means
that plane/continuous waves are unstable when perturbed, exhibiting exponential growth
rates. On the other hand, the defocusing case is modulationally stable. As such, the
instability of plane waves that obey Eq. (\ref{fl}) is rather important and as we will
see below rather interesting and somewhat different from its NLS counterpart. Indeed,
consider the continuous wave (cw) solution of Eq. (\ref{fl})
\[
u_b(t)=u_0\ue^{\ri u_0^2 \sigma t},\quad u_0\in\mathbb{R}
\]
which is perturbed as
\[
u(t, x) = [u_0 + \eps u_1(t,x)] \ue^{\ri u_0^2 \sigma t},\quad 0<\eps\ll 1.
\]
Substituting back to Eq. (\ref{fl}) and keeping terms of $O(\eps)$ gives the equation for
$u_1$,
\begin{equation}
\ri u_{1t} - \nu u_{1tx} + \gamma u_{1xx} + u_0^2\sigma(u_1+u_1^*)=0
\label{mi1}
\end{equation}
where $u_1^*$ denotes the conjugate. Eq. (\ref{mi1}) admits solutions of the form
\[
u_1(t,x)=c_1 \ue^{\ri (k x-\omega  t)} + c_2 \ue^{-\ri (k x-\omega  t)}
\]
provided the dispersion relation
\begin{equation}
(\nu^2k^2-1) \omega^2 + 2k\nu(\gamma k^2-\sigma u_0^2)\omega +
\gamma k^2 (\gamma k^2-2\sigma u_0^2)=0.
\label{disp}
\end{equation}
Notice here that in the NLS limit ($\nu=0$)
\[
\omega^2=\gamma^2k^4-2\sigma\gamma u_0^2 k^2
\]
the sign of the product $\sigma\gamma$, used in the NLS equation, is sufficient to
provide the stability conditions, namely if $\sgn(\sigma\gamma)=1$ the (focusing)
equation is unstable and when $\sgn(\sigma\gamma)=-1$ the (defocusing) equation is
stable. However, here the discriminant of Eq. (\ref{disp}) reveals that key to the
stability of the equation is the product $\sigma (-2\gamma + u_0^2 \nu^2 \sigma)$; when
positive the equation is termed stable. Hence, the stability criterion now reads:
\begin{equation}
\nu^2 > 2\sigma\gamma/u_0^2,\quad \sigma^2=1.
\label{criterion}
\end{equation}
Remarkably the same product $\sigma\gamma$ may also be used to determine stability
properties, i.e. when negative the (defocusing) system is always stable. However, the
focusing problem may now also be stable provided the above, leaving only a narrow window
of instability. Notice also that the role of the amplitude of the cw should not be
neglected. Indeed, the more its intensity increases the smaller the window of instability
becomes. Notice that the limit $\nu\rightarrow 0$ is singular as the corresponding
focusing NLS equation is always unstable.

The above analysis will prove to be very useful in what follows; it is the basis of the
solutions which will be constructed on top of this cw. Return to Eq. (\ref{fl}) and use
the Madelung transformation $u(t,x)=\rho(t,x)\exp[\ri \phi(t,x)]$, so that after
separating real and imaginary terms we get the system
\begin{eqnarray}
(1-\nu\phi_x)\rho_t + \gamma \rho\phi_{xx}+2\gamma\rho_x\phi_x +
\nu\sigma\rho^2\rho_x - \nu\left( \rho\phi_t \right)_x &=& 0 \label{real} \\
\rho(1-\nu\phi_x)\phi_t + \gamma\rho\phi_x^2 - \gamma\rho_{xx} -
\sigma(1-\nu\phi_x)\rho^3 +\nu\rho_{tx}&=& 0. \label{imag}
\end{eqnarray}
Next define the new scales
\[
T=\eps^{3/2}t, \quad X = \sqrt{\eps} (x-ct)
\]
where $0<\eps\ll 1$ a small parameter and $c$ is a travelling frame velocity to be
determined later in the analysis; this is actually the speed of sound, namely the
velocity of small-amplitude and long-wavelength waves propagating along the cw
background. Furthermore, the amplitude $\rho(t,x)$ and phase $\phi(t,x)$ are expanded in
a series of the small parameter as follows:
\begin{eqnarray*}
  \rho  &=& {\rho _0} + \eps{\rho _1} + \eps^2 {\rho _2} +  \cdots \\
  \phi  &=& \sigma\rho_0^2 t + \sqrt{\eps}\phi _1 + \eps^{3/2}{\phi _3} \cdots \\
\end{eqnarray*}
Substituting back to Eqs. (\ref{real})-(\ref{imag}) we obtain sets of equations defining
the relative fields at different orders of $\eps$. Hence at
\begin{eqnarray}
O(\eps) &:&\quad \frac{{\partial {\phi _1}}}{{\partial X}} =
- \frac{{2\sigma {\rho _0}}}{c}{\rho _1} \label{eps1}
\\
O(\eps^{3/2}) &:& \quad \frac{{{\partial ^2}{\phi _1}}}{{\partial {X^2}}}
= \frac{c}{{(\gamma  + c\nu ){\rho _0}}}\frac{{\partial {\rho _1}}}{{\partial X}}
\end{eqnarray}
the compatibility condition between the two equations (the latter is obtained by
differentiating the first with respect to $X$) yields the equation for $c$, namely:
\begin{equation}
c^2+(2\sigma\nu\rho_0^2)c+2\gamma\sigma\rho_0^2=0 \Rightarrow
c=-\nu\sigma\rho_0^2\pm\rho_0\sqrt{-2\gamma\sigma+\nu^2\rho_0^2}
\label{speed}
\end{equation}
Importantly one should notice here that the sign of $(-2\gamma\sigma+\nu^2\rho_0^2)$ is
also determined by the stability criterion (\ref{criterion}), thus suggesting that only
stable waves will propagate with real velocities and vice versa. The two signs in Eq.
(\ref{speed}) correspond to waveforms propagating with different velocities. When
$\nu=0$, i.e. in the NLS case the distinction is trivial: waveforms propagate either to
the left or to the right.

Moving to the higher orders in $\eps$ we obtain:
\begin{eqnarray}
O(\eps^2) &:& \quad \frac{{\partial {\phi _1}}}{{\partial T}} + (\gamma  + c\nu ){\left( {\frac{{\partial {\phi _1}}}{{\partial X}}} \right)^2} + \frac{{ - c + 2\nu \rho _0^2\sigma }}{{{\rho _0}}}{\rho _1}\frac{{\partial {\phi _1}}}{{\partial X}} \nonumber \\
&& \quad - \frac{{\gamma  + c\nu }}{{{\rho _0}}}\frac{{{\partial ^2}{\rho _1}}}{{\partial {X^2}}} - 3\sigma \rho _1^2= c\frac{{\partial {\phi _2}}}{{\partial X}} + 2{\rho _0}\sigma {\rho _2} \\
O(\eps^{5/2}) &:& \quad \frac{{\partial {\rho _1}}}{{\partial T}} + (\gamma  + c\nu )\left( {{\rho _1}\frac{{{\partial ^2}{\phi _1}}}{{\partial {X^2}}} + 2\frac{{\partial {\rho _1}}}{{\partial X}}\frac{{\partial {\phi _1}}}{{\partial X}}} \right) - \nu {\rho _0}\frac{{{\partial ^2}{\phi _1}}}{{\partial T\partial X}}\nonumber\\
&& \quad  + 2\nu \sigma {\rho _0}{\rho _1}\frac{{\partial {\rho _1}}}{{\partial X}} =  - {\rho _0}(\gamma  + c\nu )\frac{{{\partial ^2}{\phi _2}}}{{\partial {X^2}}} + c\frac{{\partial {\rho _2}}}{{\partial X}}
\end{eqnarray}
These equations may be uncoupled if, say, one solves for $\rho_2$ the first, substitutes
in the second and use Eq. (\ref{speed}) to eliminate $\rho_2$ and $\phi_2$. The resulting
equations is:
\begin{equation}
4\rho _0^2{(\sigma c + \nu {\rho _0})^2}\frac{{\partial {\rho _1}}}{{\partial T}}
- 2\sigma \rho _0^2{(\gamma  + c\nu )^2}\frac{{{\partial ^3}{\rho _1}}}{{\partial {X^3}}}
- 12\rho _0^3(2\gamma  + c\nu ){\rho _1}\frac{{\partial {\rho _1}}}{{\partial X}} = 0
\label{kdv}
\end{equation}
This is clearly a KdV equation whose solutions and properties may now be used for the
construction of solution of the original FL equation.

\section{Soliton solutions}

We are focusing here in the single soliton solution of Eq. (\ref{kdv}) which may be
written as
\begin{equation}
{\rho _1}(T,X) = \frac{{2{{(\gamma  + c\nu )}^2}}}{{(2\gamma  + c\nu )\sigma {\rho _0}}}\,{\eta ^2}\,
\mathrm{sech}^2\!\left[ {\eta \left( {X + \frac{{2{{(\gamma  + c\nu )}^2}}}{{c + \nu \sigma \rho _0^2}}{\eta ^2}T} \right)} \right]
\label{soliton}
\end{equation}
with corresponding phase, obtained from Eq. (\ref{eps1}),
\begin{equation}
{\phi _1}(T,X) =  - \frac{{4{{(\gamma  + c\nu )}^2}}}{{c(2\gamma  + c\nu )}}\,\eta\,
\mathrm{tanh}\! \left[ {\eta \left( {X + \frac{{2{{(\gamma  + c\nu )}^2}}}{{c + \nu \sigma \rho _0^2}}{\eta ^2}T} \right)} \right]
\label{phase}
\end{equation}
Of particular interest is the sign of the amplitude $\rho_1$, as based on the multiscale
expansion the complete solution of Eq. (\ref{fl}) is written as, to $O(\eps)$,
\[
u(t,x)=(\rho_0+\eps\rho_1)\; \ue^{\ri (\sigma\rho_0^2 t+\sqrt{\eps}\phi_1)}.
\]
As such, depending on this sign one can have intensity dips off of the constant
background $\rho_0$, corresponding to dark solitons or intensity humps on top of the
background corresponding to antidark solitons.

In what follows we fix $\rho_0=1$, with no loss of generality. To fully understand the
plethora of different solutions two cases will be considered:
\begin{enumerate}
  \item {\em The defocusing case}: Here we have $\sigma\gamma=-1$ and regardless of the
      values of $\nu$ Eq. (\ref{fl}) is always modulationally stable. Setting
      $\sigma=-\gamma=-1$ we can only obtain dark solitons (the sign of Eq.
      (\ref{soliton}) is always negative) and two propagating directions. One to the
      left with $c_L=\nu-\sqrt{\nu^2+2}$ and one to the right with
      $c_R=\nu+\sqrt{\nu^2+2}$.
  \item {\em The focusing case}: Remarkably the equation exhibits soliton solutions
      with nonzero boundary condition (provided the stability criterion
      $\nu^2>2\sigma\gamma$ is respected) even in the focusing case. Moreover, two
      different solitons exist. Indeed, as before let us set $\sigma=\gamma=1$ then
      dark solitons ($\rho_1$ is negative) exist propagating to the right with
      $c_R=-\nu-\sqrt{\nu^2-2}$ and antidark solitons ($\rho_1$ is positive) exist that
      propagate to the left with $c_L=-\nu+\sqrt{\nu^2+2}$.
\end{enumerate}

Note, finally, that in the case $\sgn(\gamma / \nu)=1$ and by replacing $u(t,x)$ by
$u(t,-x)$ a gauge transformation of the form $u\rightarrow\sqrt{\gamma/ \nu^3}\exp(\ri x/
\nu)u$ transforms Eq. (\ref{fl}) into \cite{lenells5}
\[
u_{tx}+\frac{\gamma}{\nu^3}u-\frac{2\ri \gamma}{\nu^2} u_x-\frac{\gamma}{\nu}u_{xx}
-\frac{\ri\gamma}{\nu^3}\sigma|u|^2u_x=0
\]
as such the relative analysis above also refers to the solutions of this equation as
well.

Some comments are important here. The IST for Eq. (\ref{fl}) with nonzero boundary
conditions has been presented in \cite{zhao} for the focusing case. In Ref. \cite{ling} a
plethora of single soliton are found for the coupled system. However, much like the
coupled NLS equation, a coupled system may allow for more intricate soliton pairs which
only exist in the coupled case, not the single equation case. For this they are often
termed symbiotic solitons.

\section{Conclusions}

Two integrable equations have been asymptotically connected using a multiple scales
scheme. The FL equation, derived to describe nonlinear pulse propagation in monomode
optical fibers when certain higher-order nonlinear effects are taken into account is
asymptotically reduced to a KdV equation, usually used in the theory of shallow water
waves. As such, the single soliton solution of the latter can be used to construct (small
amplitude) soliton solutions of the first. Surprisingly, the FL equation is
modulationally stable even in the focusing case, where its NLS equation counterpart is
always unstable. This allows for stable solutions to exist both in the defocusing and
focusing regimes: only dark in the first, both dark and antidark in the latter.

It is also important to mention here that this method has also been used to describe
solitons in nonlocal equations as these are formulated to describe beam propagation in
nematic liquid crystals \cite{horikis1}. In fact, in that context many intricate
solutions and relative dynamics have been revealed in coupled \cite{horikis5} and 2D
systems \cite{horikis3,horikis4} which encourage us to study coupled and 2D FL systems in
a similar fashion. We intend to do so in a future communication.


\end{document}